\documentclass[prb,twocolumn,showkeys,showpacs,amsmath,amsfonts,floatfix,superscriptaddress,nofootinbib]{revtex4-2}
\usepackage{graphicx}
\usepackage{amssymb,amsmath,amstext}
\usepackage[utf8]{inputenc}
\usepackage[T1]{fontenc}
\usepackage{url}
\usepackage{enumitem}
\usepackage{array}
\usepackage{bbold}
\usepackage{enumitem}
\usepackage{xcolor}
\usepackage{listings}
\usepackage{hyperref}

\definecolor{codegreen}{rgb}{0,0.6,0}
\definecolor{codegray}{rgb}{0.5,0.5,0.5}
\definecolor{codepurple}{rgb}{0.58,0,0.82}
\definecolor{backcolour}{rgb}{0.95,0.95,0.92}
\lstdefinestyle{mystyle}{
    backgroundcolor=\color{backcolour},   
    commentstyle=\color{codegreen},
    keywordstyle=\color{magenta},
    numberstyle=\tiny\color{codegray},
    stringstyle=\color{codepurple},
    basicstyle=\ttfamily\footnotesize,
    breakatwhitespace=false,         
    breaklines=true,                 
    captionpos=b,                    
    keepspaces=true,                 
    numbers=left,                    
    numbersep=5pt,                  
    showspaces=false,                
    showstringspaces=false,
    showtabs=false,                  
    tabsize=2
}
\begin{document}

\title{YASTN: Yet another symmetric tensor networks; \\ A Python library for abelian symmetric tensor network calculations.}

\author{Marek M. Rams}
\affiliation{Jagiellonian University, Institute of Theoretical Physics, {\L}ojasiewicza 11, 30-348 Krak\'ow, Poland}
\email{marek.rams@uj.edu.pl}
\author{Gabriela W\'ojtowicz}
\affiliation{Jagiellonian University, Institute of Theoretical Physics, {\L}ojasiewicza 11, 30-348 Krak\'ow, Poland}
\affiliation{Institut f\"ur Theoretische Physik und IQST, Albert-Einstein-Allee 11, Universit\"at Ulm, D-89081 Ulm, Germany}
\email{gabriela.wojtowicz@uni-ulm.de}
\author{Aritra Sinha}
\affiliation{Jagiellonian University, Institute of Theoretical Physics, {\L}ojasiewicza 11, 30-348 Krak\'ow, Poland}
\affiliation{Max Planck Institute for the Physics of Complex Systems, N\"{o}thnitzer Strasse 38, Dresden 01187, Germany}
\email{asinha@pks.mpg.de}
\author{Juraj Hasik}
\affiliation{Institute for Theoretical Physics and Delta Institute for Theoretical Physics, University of Amsterdam, Science Park 904, 1098 XH Amsterdam, The Netherlands}
 \affiliation{Department of Physics, University of Zurich, Winterthurerstrasse 190, 8057 Zurich, Switzerland}
\email{juraj.hasik@physik.uzh.ch}

\begin{abstract}
We present an open-source tensor network Python library for quantum many-body simulations. 
At its core is an 
abelian-symmetric tensor, implemented as a sparse block structure managed by  logical layer on top of dense multi-dimensional array backend. This serves as the basis for higher-level tensor networks algorithms, operating on matrix product states and projected entangled pair states, implemented here.
Using appropriate backend, such as PyTorch, gives direct access to automatic differentiation (AD) for cost-function gradient calculations and execution on GPUs or other supported accelerators.
We show the library performance in simulations with infinite projected entangled-pair states, such as finding the ground states with AD, or simulating thermal states of the Hubbard model via imaginary time evolution. We quantify sources of performance gains in those challenging examples allowed by utilizing symmetries.
\end{abstract}

\maketitle

\section{Introduction}

Full numerical treatment of quantum-mechanical systems is generally prohibitively expensive due to the exponential growth of Hilbert space size with the number of interacting degrees of freedom.
{\it Tensor network} (TN) techniques allow for the efficient representation and manipulation of states of such large quantum systems~\cite{orus2014practical, orus2019tensor,cirac2021matrix}. The {\it density matrix renormalization group} (DMRG) introduced by White~\cite{white1992density, white1993density} and its modern reformulation in terms of {\it matrix product states}~\cite{rommer1997class, dukelsky1998equivalence, vidal2024efficient, schuch2008entropy, schollwock2011density} (MPS), a one-dimensional tensor network, is a prime example of TN capabilities. Since their inception, MPS quickly became a reference method for addressing ground states in one dimension, and were soon followed by extensions to excited states, time evolution, and open systems forming a comprehensive framework. 

\begin{figure}[tbp]
    \centering
    \includegraphics[width=\columnwidth]{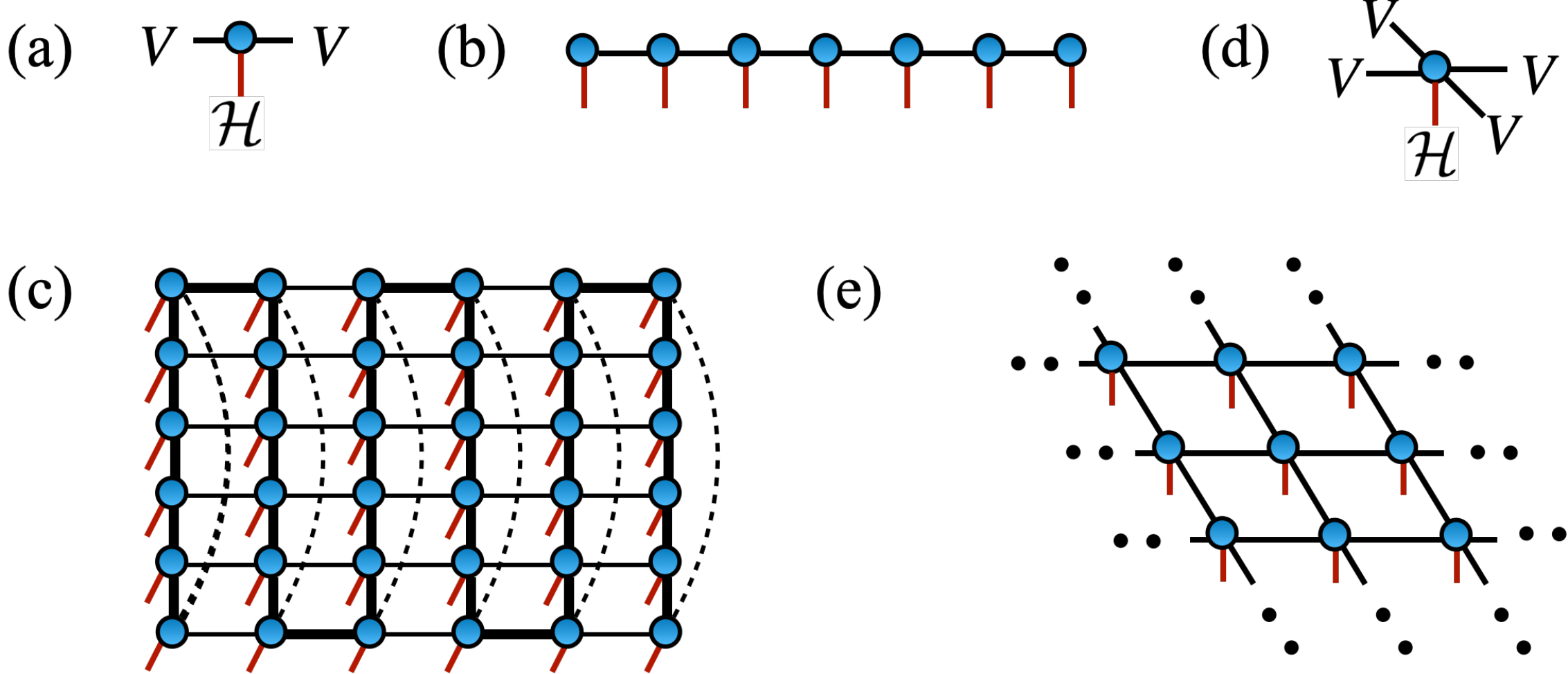}
    \caption{{\bf Tensor networks.} Diagrams depict (a) a rank-3 tensor of the MPS with left and right virtual spaces $V$ and physical space $\mathcal{H}$, (b) an MPS ansatz, (c) a MPS winding on a finite width-6 cylinder, (d) a rank-5 tensor of the PEPS, and (e) an infinite-PEPS ansatz.
    } 
    \label{fig:1}
\end{figure}

The descriptive power of tensor networks is not limited to systems in one dimension.
Despite their intrinsic 1D geometry, MPS can be readily applied to models in higher dimensions by linearly ordering the sites.
Typically, for 2D models a finite cylinder is chosen and mapped to the MPS ansatz by imposing ordering winding around the cylinder circumference as shown in Fig.~\ref{fig:1}(c). 
More natural TN geometry for two-dimensional states is assured by the {\it projected entangled-pair states}~\cite{verstraete2004renormalization, verstraete2006criticality} (PEPS)  (see Fig.~\ref{fig:1}e) and the similar for 3D states~\cite{orus2012exploring, vlaar2021simulation}. 
The TN ansätze in Fig.~\ref{fig:1}, provide a state-of-the-art numerical approach to strongly correlated systems of condensed matter. 
The computational complexity of MPS typically scales as $\mathcal O(D^3)$, and PEPS algorithms often reach $\mathcal O(D^{12})$ scaling, where bond dimension $D$ governs the size of the tensors and the overall precision of the TN approximations. 
While the scaling of PEPS seems less favorable, it is important to note that the bond dimension encodes correlations between sites. 
Under the snake-MPS ansatz, the correlations inside a column are stretched due to the row-by-row mapping, leading PEPS to reach comparable or better precision even at low bond dimensions once the cylinders become too wide.

The most effective way to mitigate the computational complexity is to leverage the symmetries present in physical systems. Two principal types of symmetries to consider are spatial and internal symmetries.
Tensor networks can be formulated directly in the thermodynamic limit by infinitely repeated pattern (a unit cell) of tensors, hence realizing translation symmetry. These are infinite-MPS (iMPS) also known as uniform MPS ~\cite{vidal2007classical} in one dimension and the infinite-PEPS (iPEPS)~\cite{jordan2008classical} in two dimensions. 
The computational complexity of iMPS/iPEPS algorithms scales linearly with the size of the unit cell.
For internal symmetries $U|\Psi\rangle=|\Psi\rangle$, we consider their common form of global symmetries, i.e., when $U=\otimes_i u_i$ with the same unitaries $u_i$ acting on each site. These can be both abelian (e.g., particle conservation) or non-abelian (e.g., SU(2)-spin). 
Crucially, such global symmetries can be implemented in TNs locally, by requiring individual tensors to transform covariantly under the action $u$ of the symmetry group~\cite{mcculloch2007density, singh2010tensor, singh2011tensor, singh2012tensor, silvi2019tensor}. These symmetric tensors take block-spare form, with original dense virtual spaces $V$ of bond dimension $D$ split into a direct sum of blocks $V=\oplus_r V_r$ with dimensions $\{D_1,\ldots,D_r\}$ each associated to irreducible representation $r$ of the considered symmetry group. 
The use of block sparsity substantially lowers computational complexity permitting large-$D$ simulations, in particular for (i)PEPS algorithms.

Here, we introduce the {\it Yet Another Symmetric Tensor Network} (YASTN) library~\cite{yastn2024}. 
YASTN is an open-source Python library with abelian-symmetric tensor as a basic type and associated linear algebra operations on such tensors. The implementation enables {\it automatic differentiation} (AD) via appropriate dense linear algebra backends,
allowing convenient variational optimization of TNs. This is particularly important for iPEPS~\cite{corboz2016grad,vanderstraeten2016grad,liao2019differentiable}, where no alternative direct energy minimization algorithms are known. This is in contrast to the (i)MPS where the DMRG 
provides efficient and robust optimization. 
YASTN thus joins a continually growing collection of tensor network software with various degree of support for symmetries and automatic differentiation such as iTensor~\cite{fishman2022itensor}, TenPy~\cite{hauschild2018efficient}, Block2~\cite{zhai2023block2}, Quantum TEA~\cite{quantumtea}, TensorNetwork~\cite{tensornetwork2019}, Cytnx~\cite{wu2023cytnx}, TeNes~\cite{tenes}, TensorKit~\cite{TensorKitJL},  Qspace~\cite{weichselbaum2024qspace},  peps-torch~\cite{pepstorch2024},  ad-peps~\cite{ad-peps}, variPEPS~\cite{naumann2023varipeps}, PEPSKit~\cite{PEPSKit}.

In the following sections, we outline the design principles of YASTN and present a set of benchmarks demonstrating the computational speed-up from abelian symmetries. We focus on variational optimization of iPEPS for SU(2)-symmetric spin-$\frac{1}{2}$ model, SU(3)-symmetric model, and observables of a Hubbard model at finite temperature simulated via imaginary-time evolution.

\section{Design principles}

\begin{figure}[htbp]
    \centering
    \includegraphics[width=0.8\columnwidth]{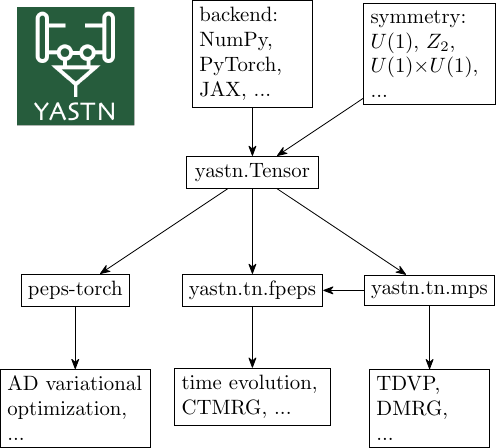}
    \caption{{\bf Schematic design of yastn package.} The core element of the package is \lstinline!yastn.Tensor!, which implements block-sparse tensor structure corresponding to a given abelian symmetry on top of a dense linear algebra backend, such as NumPy~\cite{harris2020array} or PyTorch~\cite{paszke2019pytorch}. More complex tensor networks, built on symmetric tensor, include standard but versatile MPS toolbox and 2D PEPS implementations to simulate the time evolution or ground-state variational optimization. The latter can benefit from automatic differentiation supported by some underlying backends, such as PyTorch.
    } 
    \label{fig:2}
\end{figure}

In this section, we give an overview of the structure of YASTN, presented in Fig.~\ref{fig:2}, and comment on some aspects of implementation.
The basic building block of the library is the \lstinline!yastn.Tensor! which is defined by the symmetry structure and the backend.
The symmetry structure determines allowed blocks and how to manipulate them when performing tensor algebra. 
The backend handles the execution of dense linear algebra operations and storage of tensor elements. 
These two are independent of each other. 
Symmetric tensors are used to construct TN ansätze such as MPS and PEPS, and finally define high-level algorithms that are applied to specific TN. 
For a detailed description of the library and all its functionalities, see the documentation under Ref.~\cite{yastn2024}. 

\subsection{Abelian-symmetric tensor}

Tensors are multilinear maps from products of several vector spaces
\begin{equation}
        T:\quad V^1\otimes V^2\otimes V^3\otimes... \rightarrow \mathbb{C} ,
\end{equation}
where $V^i$ is a vector space and $\otimes$ is a tensor product.
In quantum mechanical context we work with Hilbert
spaces $\mathcal{H}$ and their duals $\mathcal{H}^*$, and by choosing some bases in each of these spaces the tensors can be written out in components 
\begin{equation}
    T = \sum_{abc...ijk...} T^{abc...}_{ijk...} |i \rangle|j \rangle|k \rangle\ldots 
    \langle a |\langle b |\langle c |\ldots ,
\end{equation}
where $i,j,k\dots$ are indices of bases in $\mathcal H$ spaces, $a,b,c\dots$ in $\mathcal H^*$ spaces, and $T^{abc...}_{ijk...}$ is the corresponding tensor element. 
The action of the element $g$ of abelian group $G$ on tensor $T$
can be represented in a proper basis by diagonal matrices $U^{i}(g)$ transforming the tensor elements
\begin{equation}
\begin{split}
    (g\circ T)^{ab...}_{ij...} = &\sum_{a'b'...i'j'...} T^{a'b'...}_{i'j'...} {[U^{1}(g)}]^{i'}_{i} {[U^{2}(g)}]^{j'}_{j}\ldots \\ &\quad\times [U^{m}(g)^*]^{a}_{a'} [U^{m+1}(g)^*]^{b}_{b'} \ldots.
\end{split}
\end{equation}
The matrix elements of $U^{i}(g)$ are
\begin{equation}
    {[U^{i}(g)]}^{j}_{j'}=\delta_{jj'} \, \textrm{exp}({-i\theta_g t^{[i]}_j}) ,
\end{equation}
forming a diagonal matrix of complex phases defined by integer charges $t^{[i]}_j$, with angle $\theta_g \in [0,2\pi)$ which depends on $g \in G$, and $\delta_{jj'}$ being
Kronecker delta. 
Therefore, under the action of $g\in G$ each tensor element simply acquires a phase given by the sum of charges
\begin{equation}
\begin{split}
    (g\circ T)^{ab...}_{ij...} = T^{ab...}_{ij...} \textrm{exp}[-i\theta_g(&t^{[1]}_i + t^{[2]}_j+\ldots\\
    &- t^{[m]}_a - t^{[m+1]}_b - \ldots)].
\end{split}
\end{equation}
This form of the transformation gives a simple selection rule, a charge conservation, on the elements of symmetric tensors
\begin{equation}
\label{eq:elemchargecon}
    t^{[1]}_i+t^{[2]}_j+...-t^{[m]}_a-t^{[m+1]}_b-... = n .
\end{equation}
The charge of each non-zero element $T^{ab...}_{ij...}$ of a symmetric tensor must be $n$.
In the case of $n=0$, such tensor is invariant (unchanged) under the action of the symmetry.
Otherwise, it transforms covariantly as all its elements are altered by the same complex phase $\textrm{exp}(-i\theta_g n)$.
The charges $t^{[i]}_j$ and $n$ and the precise form of their addition depend on the abelian group. For elementary abelian groups such as $Z_n$ or $U(1)$ the individual  charges $t^{[i]}_j$ are elements of $\mathbb{Z}_n$ or $\mathbb{Z}$ respectively, while for direct products of abelian groups, they become vectors in corresponding product of $\mathbb{Z}_n$'s and $\mathbb{Z}$'s.

By ordering the basis elements in each Hilbert space by their charge, the tensor $T^{ab...}_{ij...}$ naturally attains a block-sparse structure 
which is central to the computational advantage offered by abelian-symmetric tensor network algorithms.

At the core of YASTN is the implementation of symmetric tensor \lstinline!yastn.Tensor!, as outlined in Refs.~\cite{singh2010tensor,singh2011tensor,mcculloch2007density}. It is defined jointly by symmetry (block) structure data and tensor elements of existing blocks. First, we define a vector space with a charge structure,  a \lstinline!yastn.Leg!, determined by a signature $s=\pm 1$ (distinguishing between $\mathcal{H}$ and dual $\mathcal{H}^*$), its charge sectors $\bf t$, and their corresponding dimensions $\bf D$, now in boldface to underline their vector nature,
\begin{equation}
    V(s,{\bf t}, {\bf D})=\oplus_{\rho \in {\bf t}} \mathbb{C}^{{\bf D}_\rho},
\end{equation}
where $\rho$ now enumerates different charges instead of basis 
\footnote{The conjugation of the leg, i.e., mapping space $\mathcal H$ to its dual space  $\mathcal H^*$, is equivalent to flip of the signature and complex conjugation of elements.}. 
This space is a direct sum of simple spaces $\mathbb{C}^{{\bf D}_\rho}$, dubbed charge sectors. The effective dimension of such space is the sum of dimensions of individual charge sectors
\begin{equation}
    D = \sum_\rho {\bf D}_\rho.
\end{equation}
In the remainder of the text, we will refer to $D$ as the bond dimension, 
when discussing scaling of computational complexity or memory requirements of TN algorithms with symmetric tensors.

The abelian symmetric tensor of rank-$N$ is specified by the product of $N$ such vector spaces
\begin{equation}
    T: \otimes^N_{i=1} V(s^{[i]}, {\bf t}^{[i]}, {\bf D}^{[i]}) \rightarrow \mathbb{C},
\end{equation}
The following is an example creating a random $U(1)$-symmetric tensor with total charge
~\footnote{One can always define tensors with an extra dummy leg $V(-1,(n,),(1,))$, having a single charge sector of a unit dimension, making it invariant under symmetry transformations.} 
$n=0$ with specified legs:
\begin{lstlisting}[language=Python]
import yastn
from yastn.backend import backend_np
from yastn.sym import sym_U1

u1 = yastn.make_config(sym=sym_U1, backend=backend_np)
l = yastn.Leg(u1, s=1, t=(-1,1), D=(1,1))
lc = l.conj()
H = yastn.rand(u1, legs=[l,l,lc,lc], n=0)
\end{lstlisting}
which is, for example, compatible with a Hamiltonian $H=\vec{S}_1\cdot\vec{S}_2$ of two spin-$\frac{1}{2}$ degrees of freedom.
The configuration created by \lstinline!yastn.make_config! specifies the symmetry, e.g., \lstinline!yastn.sym.sym_U1! for the $U(1)$ in the example,
and dense linear algebra backend (see below), e.g., \lstinline!yastn.backend.backend_np! for NumPy. 
The covariant transformation property of $T$ is imposed by the charge conservation 
of non-zero blocks. Any block of tensor $T$ can be identified by selecting a charge sector $\rho_i \in {\bf t}^{[i]}$ on each of the legs, i.e., an $N$-tuple of charges 
$(\rho_0,\rho_1,...,\rho_N)$. All non-zero blocks must satisfy
\begin{equation}
    \sum_{i=1}^N s^{[i]} \rho_i = n,
\end{equation}
which is the block-sparse version of element-wise charge conservation rule of Eq.~\ref{eq:elemchargecon}. 

Finally, we remark on the storage of tensor elements. In YASTN, the block data is initialized \textit{lazily}. The storage is allocated only for the blocks which have been assigned a non-zero value, i.e., blocks allowed by the charge conservation but not assigned any value are not stored. All allocated blocks are serialized together in a 1D array.

\subsection{Fusion and contractions}

The key operations on symmetric tensors, necessary for manipulating tensor networks, are tensor reshape and permutation, commonly dubbed {\it fusion} in this context, and tensor contractions. 
Fusion resolves the tensor product of several spaces as a new space, i.e., fusion of legs into a new leg. Unlike reshaping of the dense tensor, the shape cannot be chosen freely. Instead, it is determined by the structure of fused spaces.
In particular, fusion orders and accumulates tensor products of charge sectors on selected legs into new charge sectors under the joint leg
\begin{equation}
    V(s^{[i]}, {\bf t}^{[i]}, {\bf D}^{[i]}) \otimes V(s^{[j]}, {\bf t}^{[j]}, {\bf D}^{[j]}) \rightarrow V(s^{[r]}, {\bf t}^{[r]}, {\bf D}^{[r]}),
\end{equation}
with new charge sectors ${\bf t}^{[r]}$ given by the unique combinations of charges ${\bf t}^{[i]} \otimes {\bf t}^{[j]}$
\begin{equation}
    {\bf t}^{[r]} := \{ \nu = s^{[r]}(s^{[i]}\rho + s^{[j]}\rho') : \rho \in {\bf t}^{[i]}, \rho' \in  {\bf t}^{[j]}\}.
\end{equation}
The dimension of new charge sector $\nu \in {\bf t}^{[r]}$ is 
\footnote{Following a {\it lazy} approach adopted in YASTN, a new fused leg contains only charges for which some non-zero tensor block exists. As such, fusion in YASTN is always done in the context of particular tensor.}
\begin{equation}
    {\bf D}^{[r]}_\nu = \sum_{\substack{\rho,\rho' \\ \nu = s^{[r]}(s^{[i]}\rho + s^{[j]}\rho')}} {\bf D}^{[i]}_\rho {\bf D}^{[j]}_{\rho'}.
\end{equation}
The fusion and un-fusion calls are demonstrated below on previously constructed rank-4 tensor $H$, first fusing pairs of legs resulting in a matrix form 
\begin{lstlisting}[language=Python]
H_mat = H.fuse_legs(axes=((0,1), (2,3)))
H = H_mat.unfuse_legs(axes=(0,1))
\end{lstlisting}
In the example above, the YASTN first computes the structure of the resulting tensor with fused leg, i.e., the tuple $(s^{[r]}, {\bf t}^{[r]}, {\bf D}^{[r]})$ and a set of dense linear algebra jobs (permutes, reshapes, and copies) to be executed by the backend to populate the new 1D storage array with serialized blocks.
The resulting tensor records the original structure and hence, the fusion can be reverted to restore the original structure.

The tensor contraction of symmetric tensors is realized by a commonly adopted workflow. First, the tensors are fused into matrices
\footnote{For valid contraction, the structures of the contracted legs must be compatible, including the origins of any fused leg. YASTN automatically resolves a situation when some charge in the fusion history is missing in a to-be-contracted fused leg but is present in its contraction partner, utilizing information on the tensor's fusion history stored in each tensor.}, then multiplied along the contracted legs, and finally unfused to obtain the desired form: 
\begin{equation}
\begin{split}
    \sum_{\{x\}} A_{\{i\} \cup \{x\}} B_{\{j\} \cup \{x\}} &\overset{\text{fuse}} \longrightarrow 
    \sum_X A_{IX} B_{XJ} = \\ 
    C_{IJ} &\overset{\text{unfuse}} \longrightarrow C_{\{i\}\{j\}}
\end{split}
\end{equation}
where $\{x\}$ is a set of common legs that become fused into single leg $X$, and original legs $\{i\}$ and $\{j\}$ are restored from the fused $I$ and $J$ to obtain the final tensor. Here, we first show an example call for pairwise tensor contraction, and second, an equivalent given in terms of explicit operations
\begin{lstlisting}[language=Python]
H2 = yastn.tensordot(H, H, axes=((2,3), (0,1)))
H2 = (H_mat @ H_mat).unfuse_legs(axes=(0,1))
\end{lstlisting}
For both fusion and multiplication, the YASTN first precomputes what are the non--zero blocks so the backend performs only the relevant operations. 
Using these elementary operations, contractions of more general networks are supported through convenience functions, such as the \lstinline!einsum! and \lstinline!ncon!~\cite{pfeifer2015ncon} functions (in our case differing only by syntax).

\subsection{Tensor network algorithms}

The symmetric tensor serves as a foundation for higher-level tensor network structures and algorithms.
Here, the YASTN comes with MPS and PEPS modules. The MPS module supports finite-size MPS with the implementation of a range of standard algorithms, including DMRG for ground-state optimization, TDVP~\cite{haegeman2016unyfying} for time evolution, and the overlap maximization~\cite{verstraete2008matrix} against a general target, i.e., MPS, sum of MPSs, or sum of MPO-MPS products. This is complemented by a versatile high-level (Hamiltonian) MPO generator. 
The MPS module provides subroutines for some PEPS methods, e.g., it was utilized in~\cite{king2024computational} for boundary MPS contraction and long-range correlations calculation in a finite PEPS defined on a cylinder.
At the same time, it is a versatile computational toolbox on its own. For instance, it has been employed in simulations of Lindbladian dynamics in the context of fermionic quantum transport~\cite{wojtowicz2023accumulative}, where the $U(1)$ symmetry reflects a lack of correlations between different particle-number sectors of a density matrix.

The PEPS module features the implementation of fermionic PEPS, dubbed fPEPS (which also allows simulations of systems without fermionic statistics). It covers both finite PEPS defined on a square grid and its infinite versions for translationally invariant (over a unit-cell) systems in the thermodynamic limit. It supports a range of time-evolution algorithms, starting with neighborhood tensor update (NTU) scheme~\cite{dziarmaga2021time, dziarmaga2022simulation}, its refinement to a family of larger environmental clusters~\cite{king2024computational}, ending on a full-update type of schemes~\cite{jordan2008classical,czarnik2019time}. It is viable for imaginary-time evolution, e.g., in the context of finite temperature simulation of density matrix purification~\cite{sinha2022finite-temperature} or minimally-entangled typical thermal states~\cite{sinha2024efficient}, and real-time simulations, e.g., of pure state quench-dynamics in disordered spin systems~\cite{king2024computational}.

\section{Examples}
To demonstrate the use and the versatility of YASTN we present three end-to-end numerical examples centered on iPEPS. %
We show computational speed-up and reduced memory footprint obtained with YASTN by utilizing abelian symmetries for the following examples:
\begin{enumerate}
\item Sec.~\ref{sec:hafm}: variational optimization of iPEPS with $D\leq8$ for antiferromagnetic spin-$\frac{1}{2}$ model on a square lattice using $U(1)$ symmetry,
\item Sec.~\ref{sec:su3kag}: variational optimization of iPEPS with $D\leq13$ for SU(3) model on Kagome lattice using $U(1){\times}U(1)$ symmetry,
\item Sec.~\ref{sec:thdhubbard}: observables of thermal iPEPS of Hubbard model at finite temperature using $Z_2$, $U(1)$, and $U(1){\times}U(1)$ symmetry, with $D$ up to $36$ for the latter.
\end{enumerate}

In Sec.~\ref{sec:hafm} and Sec.~\ref{sec:su3kag} we optimize iPEPS for SU(2) model on square and SU(3) model on Kagome lattices respectively. 
First, given an iPEPS generated by a set of tensors $\vec{a}=\{a,b,\ldots\}$, we compute an approximate environment tensors $\vec{E}(\vec{a})$ (specified below) with the precision governed by the environment dimension $\chi$. 
Then, environment $\vec{E}$ and tensors $\vec{a}$ are combined to evaluate the energy per site $e$ of the Hamiltonian. 
Finally, the reverse mode of AD (i.e., backpropagation) is invoked to compute the gradient $\partial e/\partial \vec{a}$. 
The most computationally intensive stage is the construction of the environments, scaling as the cube of $D^2\chi$, which assuming the necessary $\chi\propto D^2$ gives the overall complexity $\mathcal O(D^{12})$, where $D$ is the bond dimension of iPEPS tensors. 
We use iPEPS optimization implemented in peps-torch~\cite{pepstorch2024}, here operating on YASTN's symmetric tensors.  

\begin{figure}[htbp]
    \centering
    \includegraphics[width=\columnwidth]{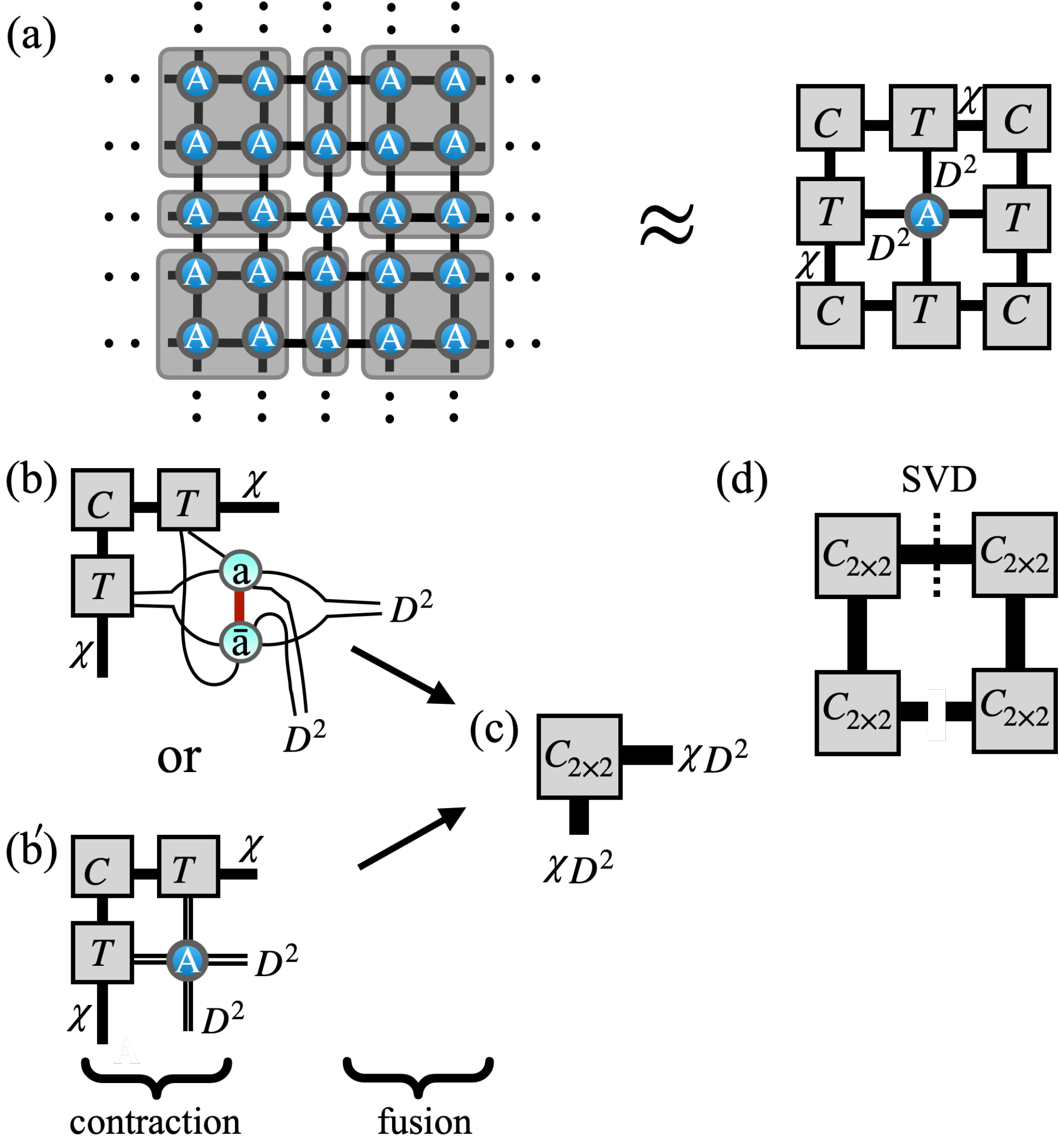}
    \caption{{\bf Corner transfer matrix iteration}. In (a), CTM approximates parts of an infinite tensor network by a set of finite environment tensors characterized by environment bond dimension $\chi$. We depict elements of the CTM algorithm step (iteration) that dominate the computational effort shown in Figs.\ref{fig:4}--\ref{fig:6}. Panels (b), (b'), and (c) depict the construction of an enlarged corner combining CTM environment tensors (rectangular) with PEPS tensors (circle). Panel (d) shows an SVD decomposition of a product of four enlarged corners that is then used to construct the CTM projections from enlarged virtual spaces.} 
    \label{fig:3}
\end{figure}

For the presented examples, we employ the {\it corner transfer matrix} (CTM) algorithm~\cite{nishino1996corner, orus2009simulation, corboz2014competing} to compute the environments. CTM approximates environments $\vec{E}=\{C,T\}$ of iPEPS by a set of $\chi\times\chi$ corner matrices $C$ and $\chi\times D^2\times\chi$ transfer tensors $T$, as shown in Fig.~\ref{fig:3}(a).
Alternatively, one can use boundary MPS methods~\cite{jordan2008classical, verstraete2008matrix, fishman2017faster}. 
The computational complexity of CTM arises from two sources, see Fig.~\ref{fig:3}, tensor contractions and {\it singular value decomposition} (SVD) when computing low-rank approximations, both scaling as $\mathcal O(D^{12})$. 
In practice, for simulations without symmetries the SVD gives a substantially greater contribution due to a higher scaling prefactor and a poor speed-up offered by multithreading or GPU acceleration compared to tensor contractions. However, for symmetric iPEPS the situation becomes more nuanced as we demonstrate in the examples below.

In Sec.~\ref{sec:thdhubbard}, the purification techniques are applied to compute thermal expectation values for the Fermi-Hubbard model on a $2$D square lattice. The ancilla trick is used to effectively transform the thermal density matrix into a purified wavefunction and then imaginary time evolution is performed to reach the target temperature. We adopt the NTU algorithm to optimize the time-evolution, with computational scaling of $\mathcal O(D^8)$ dominated by tensor contractions. We focus our example on the final calculation of the expectation values using CTM with $\chi=5 D$, translating to $\mathcal O(D^9)$ scaling. We quantify the sources of advantage offered by incorporation of higher symmetries.

\subsection{Heisenberg antiferromagnet with anisotropy}
\label{sec:hafm}

We revisit a Heisenberg model with anisotropy in the couplings describing a system of coupled spin-$\frac{1}{2}$ ladders
\begin{equation}\label{eq:ham_ex1}
{\cal H} = J \sum_{R} {\bf S}_{R} \cdot {\bf S}_{R+\hat{x}} + \sum_{R} J_{R} {\bf S}_{R} \cdot {\bf S}_{R+\hat{y}},
\end{equation}
where ${\bf S}_{R}=(S^x_{R},S^y_{R},S^z_{R})$ is the $S=\frac{1}{2}$ operator on the site $R=(x,y)$ of a square lattice, $\hat{x}$ and ${\hat{y}}$ are unit vectors in $x$ 
and $y$ direction, and $J_{R}=J$ or $J_{R}=\alpha J$, depending on the parity of $y$. By varying $\alpha$, this model interpolates between the Heisenberg model 
on the square lattice at $\alpha=1$ and a system of decoupled two-leg ladders at $\alpha=0$. The model was previously addressed by iPEPS in Ref.~\cite{hasik2022symmetric}. 
We adopt the same description that uses four non-equivalent tensors $\vec{a}=\{a,b,c,d\}$ arranged in a $2{\times}2$ unit cell~\footnote{A more efficient description might generate all tensors in $2{\times}2$ unit cell from a single parent tensor $a$ by use of unitary $-i\sigma^y$ acting on physical index and/or permutation of virtual indices generated by the reflection along the $x$-axis. Nevertheless, such parametrization would not change CTM complexity.}. 

\begin{figure}[t!]
    \centering
    \includegraphics[width=\columnwidth]{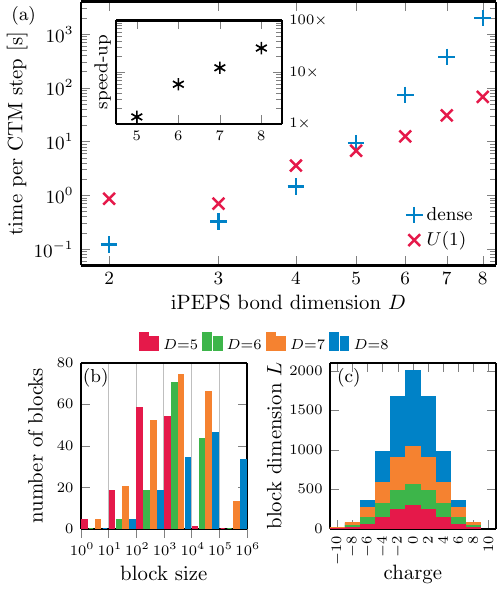}
    \caption{{\bf Use case 1: optimization of $U(1)$-symmetric iPEPS for model of spin-$\frac{1}{2}$ coupled ladders.} In (a), we show the scaling of the wall time per CTM step (in seconds) for the entire gradient optimization step of iPEPS. The bond dimension of the environment is $\chi=2D^2$. The inset shows the relative speed-up compared to an implementation without symmetry.
    In (b), we show a distribution of blocks of an enlarged corner by their size (number of elements) before fusion to a block-diagonal matrix as shown in Fig.~\ref{fig:3}(c). In (c), we show the sizes of $L\times L$ blocks after the fusion. 
    }
    \label{fig:4}
\end{figure}

The ground states possess $U(1)$-symmetry corresponding to the conservation of $S^z$ which we leverage to work with $U(1)$-symmetric iPEPS. 
The results, summarized in Fig.~\ref{fig:4}, show a rapidly growing computational advantage of symmetric iPEPS for bond dimensions $D>4$.
While at $D=4$ the overhead due to the block-sparse logic is still significant, at the largest bond dimension considered, $D=8$, a $30$-fold speed-up is observed. 
In practical terms, the convergence of the CTM towards the desired precision, here measured by the error on the energy  per site becoming lower than $\epsilon<10^{-8}$, typically requires $\mathcal O(10)$ iterations, thus without $U(1)$ symmetry a single optimization step would already take hours. 
Details of the block-sparse structure and its impact on the CTM are visualized in Fig.~\ref{fig:4}(b,c). 
At the largest bond dimension, $D=8$, 
the fusion of enlarged corner into a block-diagonal matrix requires processing of roughly $\mathcal O(100)$
blocks by performing dense permutes, reshapes, and copies, with the largest block having $\mathcal O(10^5)$ elements. The cost of subsequent SVD is dominated by the largest block(s) of fused enlarged corner, which are $L\times L$ matrices with $L=1500\sim2000$. 
As a result, the computational time contributions are shared between the SVD and tensor contractions with fusions roughly as 3:2, with SVD being the dominant factor.
 
\subsection{SU(3) model on Kagome lattice}
\label{sec:su3kag}

We consider an SU(3)-symmetric model on Kagome lattice, analyzed recently in Ref.~\cite{xu2023phase}, where each site holds a single degree of freedom from the fundamental representation $\mathbf{3}$ of SU(3) group spanned by three states $\{|\alpha\rangle,|\beta\rangle,|\gamma\rangle\}$. The Hamiltonian reads 
\begin{eqnarray}
	\label{eq:modelsu3}
	H &=& J\sum_{\langle i,j\rangle}P_{ij} + \sum_{\triangle ijk}(KP_{ijk} + h.c.),
\end{eqnarray}
where $P_{ij}$ is a permutation of local states on nearest-neighbour bonds such that $P_{ij}|\alpha\rangle_i |\beta\rangle_j = |\beta\rangle_i |\alpha\rangle_j$, 
$P_{ijk}$ is a clockwise permutation of local states on nearest-neighbour triangles such that  $P_{ijk}|\alpha\rangle_i |\beta\rangle_j |\gamma\rangle_k= |\gamma\rangle_i |\alpha\rangle_j |\beta\rangle_k$, with fixed choice of the orientation of triangles, 
and $J$ and $K$ are real- and complex-valued couplings respectively. 

\begin{figure}[t!]
    \centering
    \includegraphics[width=\columnwidth]{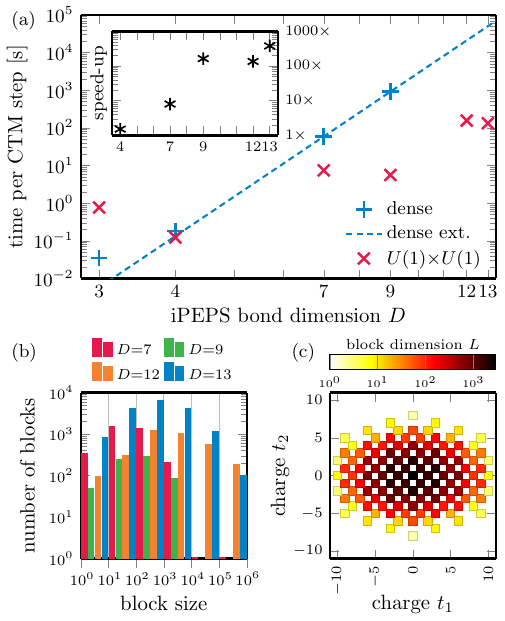}
    \caption{{\bf Use case 2: optimization of $U(1){\times}U(1)$-symmetric iPEPS for SU(3) Kagome model.}
    In (a),  scaling of the wall time per CTM step (in seconds) for the entire gradient optimization step of iPEPS. The bond dimension of
    the environment is $\chi = D^2$. The inset shows the relative speed-up compared to an implementation without symmetry, with $D=12$ and $13$ simulation wall times estimated from the extrapolation (blue dashed line). 
    In (b), a distribution of blocks of an enlarged corner by their size (number of elements) before the fusion to a matrix.
    In (c), $L\times L$ blocks of the block-diagonal enlarged corner after fusion. We plot them as a heatmap, with different $U(1)$ charges on x- and y-axes.
    }
    \label{fig:5}
\end{figure}

In this section, we demonstrate an advantage of $U(1){\times}U(1)$-symmetric iPEPS, utilizing maximal abelian subgroup of SU(3), over implementation without symmetries~\footnote{In this example, besides iPEPS, one can also use different ways to construct two-dimensional TN ansatz on Kagome lattice, i.e., infinite projected simplex states (iPESS), however, the computational complexity $\mathcal O(D^{12})$, attributable to CTM, remains unchanged.}. 
To compute CTM environments on the Kagome lattice, we coarse-grain three sites on each down-pointing triangle into a single tensor resulting in an effective square lattice. The local Hilbert space dimension thus grows to $3^3=27$, which makes the optimizations memory-intensive.
In Fig.~\ref{fig:5}, we demonstrate the dramatic speed-up achieved by utilizing $U(1){\times}U(1)$ symmetry. 
For $D=9$ iPEPS, a single gradient step is already accelerated by more than a factor of $100$. For larger bond dimensions, the simulations without symmetries become prohibitive and we estimate the speed-up based on extrapolation of the scaling at smaller bond dimensions.

In contrast to the example in Sec.~\ref{sec:hafm} utilizing the $U(1)$-symmetry, the speed-up in this case is not monotonic in $D$. 
This happens due to the varying structure of the iPEPS tensors, i.e., the allowed symmetry sectors and their sizes. In Fig.~\ref{fig:5}(b,c) we illustrate the block structure of enlarged corners before and after the fusion to a block-diagonal matrix. 
Generally, for larger groups the number of blocks of enlarged corner before fusion is substantially higher. Even at $D=7$, the total number of blocks is already more than $3000$ whereas for $U(1)$-symmetric enlarged corner in Sec.~\ref{sec:hafm} it was below $300$. For $D=13$ ansatz, the fusion of enlarged corner into a block-diagonal matrix requires processing of more than $16000$ blocks, with more than half of them being small in size, having roughly $\mathcal O(10^3)$ elements or less. This granularity defines the bottleneck of the simulations. For $D=12$ and $D=13$ the ratios between the computational time of SVD and contractions including fusion to block-diagonal enlarged corners are 3:4 and 1:10
respectively. Overall, the $U(1){\times}U(1)$ simulations become dominated by fusion, with SVD being subleading. The precise speed-up rests on the sizes of blocks, such as here, where $D=12$ has a slightly higher proportion of largest blocks compared to $D=13$. 

\begin{figure}[t!]
    \centering
    \includegraphics[width=\columnwidth]{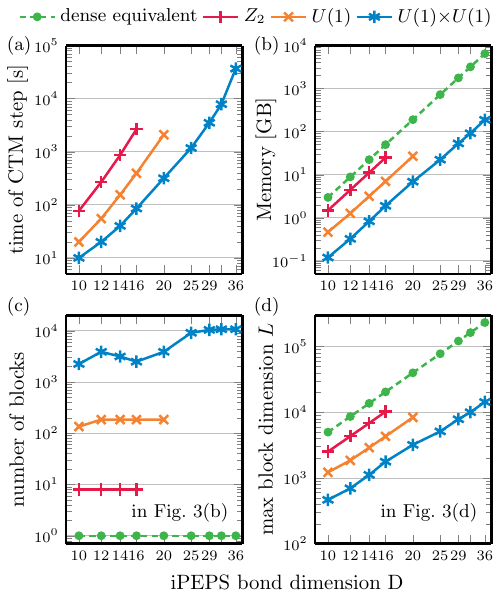}
    \caption{{\bf Use case 3: expectation values from CTM environments in fermionic iPEPS for finite-temperature Hubbard model.} The bond dimension of the environment is $\chi=5D$, which is sufficient here to converge the expectation values. 
    We show, in (a), the wall time per CTM step and, in (b), the data size (memory requirement) of the biggest intermediate tensor appearing while building enlarged CTM corners in Fig.~\ref{fig:3}(b).
    Panel (c) shows the number of blocks in an enlarged corner before the fusion in Fig.~\ref{fig:3}(c), and panel (d) is the size of the largest $L\times L$ block for SVD in Fig.~\ref{fig:3}(d).
    }
    \label{fig:6}
\end{figure}

\subsection{2D Fermi-Hubbard model on a square lattice}
\label{sec:thdhubbard}

We consider a two-dimensional Fermi-Hubbard model (FHM) with on-site repulsion as studied in Ref.~\cite{sinha2022finite-temperature}. 
The Hamiltonian reads
\begin{equation}\label{eq:fhm}
H = -t \sum_{\langle i,j \rangle, \sigma} \left(c^\dagger_{i\sigma} c_{j\sigma} + \text{h.c.}\right) + U \sum_i n_{i\uparrow} n_{i\downarrow} - \mu \sum_i n_i ,
\end{equation}
where $c^\dagger_{i\sigma}$ and $c_{i\sigma}$ are the creation and annihilation operators for an electron with spin $\sigma$ at site $i$, $n_{i\sigma} = c^\dagger_{i\sigma} c_{i\sigma}$ is a corresponding number operator, $t$ is the hopping amplitude, $U$ is the on-site Coulomb repulsion strength, and $\mu$ is the chemical potential.

The iPEPS ansatz employs a checkerboard lattice with a 2-site unit cell. The thermal state for inverse temperature $\beta$ is obtained by evolving the infinite-temperature purification $|\psi(0)\rangle$ under the propagator $U(\beta) = e^{-\frac{\beta}{2}H}$. 
The initial purification $|\psi(0)\rangle = \prod_{j} \prod_{m=\uparrow,\downarrow} \frac{1}{\sqrt{2}} \sum_{s_{m_j} = a_{m_j} = 0,1} |s_{m_j} a_{m_j}\rangle$ is a product of maximally entangled pairs between each physical site and its corresponding ancilla, translating to local Hilbert spaces of dimension $4^2=16$. In the examples below, we run the imaginary time evolution employing the NTU scheme targeting $\beta=2$. 

In YASTN, the fermionic exchange order is implemented following the scheme of Refs.~\cite{corboz2009fermionic, corboz2010simulation} by projecting the lattice ansatz onto a plane, imposing a canonical fermionic order, and applying swap gates on every line crossing that arises; see Fig.~\ref{fig:3}(b). Swap gate introduces sign changes for blocks with charges of odd parity on both swapped legs. This makes $Z_2$ a minimal symmetry needed for fermionic system simulations. The Hamiltonian in Eq.~\eqref{eq:fhm} preserves the number of particles per spin direction, which allows us to implement the model under $U(1){\times}U(1)$ symmetry as the highest abelian symmetry. 

The expectation values of the thermal state at final $\beta=2$ are calculated using the CTM. Fig.~\ref{fig:6} demonstrates an advantage of symmetric tensors by comparing $Z_2$ (parity; minimal requirement for fermionic statistics), $U(1)$ (total charge conservation), and $U(1){\times}U(1)$ (total charge conservation for each spin) symmetries. We also show equivalent values for the corresponding dense tensors with no symmetry. Here, we choose the environmental bond dimension $\chi=5D$. 

Fig.~\ref{fig:6}(a) presents the computational wall time for one CTM step as a function of the iPEPS bond dimension for all the tested symmetries, with systematic improvement offered by higher symmetries. Fig.~\ref{fig:6}(b) highlights the memory usage bottleneck, showcasing the size of the largest object formed during the CTM iteration, i.e., an intermediate step of the contraction in Fig.~\ref{fig:3}(b); this tensor has to be later fused, and there are other tensors in the memory, so the memory peak is roughly two times higher. It illustrates that those simulations are ultimately memory-limited. Employing $U(1){\times}U(1)$-symmetry offers a systematic $30$-fold memory gain as compared to tensors with no symmetries, which ultimately allows for successful simulations up to $D=36$.

Following the previous examples, in Fig.~\ref{fig:6}(c), we present the number of blocks processed during the fusion that forms enlarged corners in Fig.~\ref{fig:3}(c). A particular challenge for $U(1){\times}U(1)$ case is that the number of blocks can exceed 10000. Nevertheless, the SVD is a dominant factor taking at least half of the simulation time for $D\ge25$ in our numerical experiments. In Fig.~\ref{fig:6}(d), we show the (sectorial) bond dimension of the largest block decomposed in Fig.~\ref{fig:3}(d), that is $\mathcal O(10^4)$ for the largest bond dimension achieved with each employed symmetry. The $U(1){\times}U(1)$-symmetry offers here $15$-fold improvement as compared to a setup with no symmetries involved.

\section{Future outlook}

Tensor networks are becoming increasingly popular tool for numerical treatment of quantum systems, ranging from ground state simulations of condensed-matter systems to simulation of quantum circuits. The landscape of associated software is continuously growing. For 1D and quasi-1D geometries, well-established and mature packages offer a rich set of MPS algorithms covering direct energy minimization, (imaginary) time evolution, and much more. For two-dimensional geometries, predominantly targeted by iPEPS, the field remains nascent.

Here, we have introduced YASTN, a python-based TN library with strong emphasis on simulation of two-dimensional systems by iPEPS, motivated by the need for both abelian symmetries and automatic differentiation.   
By design, the dense linear algebra and the AD engine are provided by different backends, allowing for implementation-specific optimizations. YASTN, with its rich set of examples covering ground state simulations of various 2D spin lattice models (through peps-torch) and finite-temperature simulations of 2D Hubbard model, thus joins similar efforts by VariPEPS~\cite{naumann2023varipeps}, PEPSkit~\cite{PEPSKit}, ad-peps~\cite{ad-peps}, and peps-torch~\cite{pepstorch2024} together lowering the barrier for entry.

The wide separation between the high-level description of iPEPS algorithms and their fast execution, optimized down to low-level dense linear algebra, especially for symmetric tensors, remains a challenge. Unlike MPS simulations, iPEPS contraction algorithms for computing environments and evaluation of observables involve a more diverse set of tensor contractions, varying in ranks and block sparsity patterns.
Furthermore, flexible deployment and the ability to leverage heterogenous clusters, accounting for the iPEPS-specific block sparsity, is vital for addressing the sharp $\mathcal O(D^{8}\sim D^{12})$ (albeit polynomial) scaling with the bond dimension, which is the key resource governing the precision of iPEPS. 
These challenges thus call for further development.

\acknowledgements
We thank Philippe Corboz, Piotr Czarnik, Jacek Dziarmaga, Boris Ponsioen, Yintai Zhang for inspiring discussions that were invaluable in the development of this package.
We acknowledge the funding by the National Science Center (NCN), Poland, under projects 2019/35/B/ST3/01028 (A.S.), 2020/38/E/ST3/00150 (G.W.), and project 2021/03/Y/ST2/00184 within the QuantERA II Programme that has received funding from the European Union Horizon 2020 research and innovation programme under Grant Agreement No. 101017733 (M.M.R.). J.H. acknowledges support from the European Research Council (ERC) under the European Union's Horizon 2020 research and innovation programme (grant agreement No 101001604) and from the Swiss National Science Foundation through a Consolidator Grant (iTQC, TMCG-2\_213805).

\bibliography{refs.bib} 

\end{document}